\documentclass[12pt,a4paper]{revtex4}
\usepackage{amsmath}
\usepackage{amssymb}
\usepackage{bm}
\usepackage{graphicx}
\usepackage{wrapfig}
\usepackage{wrapft}
\begin{document}

\title{%
Study of the low-temperature behavior of a disordered antiferromagnet with random fields
by the parallel-tempering method%
}

\author{
Vladimir V. \textsc{Prudnikov}\footnote{E-mail: prudnikv@univer.omsk.su},
Andrey N. \textsc{Vakilov},
Evgenii L. \textsc{Filikanov},%
}

\affiliation{%
Dept. of Theoretical Physics, Omsk State University, Omsk 644077, Russia
}

\begin{abstract}
The parallel-tempering method has been applied to numerically study the thermodynamic behavior
of a three-dimensional disordered antiferromagnetic Ising model with random fields at spin concentrations corresponding
to regions of both weak and strong structural disorder. An analysis of the low-temperature behavior
of the model convincingly shows that in the case of a weakly disordered samples there is realized an antiferromagnetic
ordered state, while in the region of strong structural disorder the effects of random magnetic fields
lead to the realization of a new phase state of the system with a complex domain structure consisting of antiferromagnetic
and ferromagnetic domains separated by regions of a spin-glass phase and characterized by a spinglass
ground state.
\end{abstract}

\maketitle

\section{Introduction\label{sec:1}}

Studying the critical behavior of disordered systems with quenched
structural defects is of a great theoretical and experimental
interest since the majority of real solids contain quenched
structural defects whose presence influences their thermodynamic
characteristics and, in particular, substantially affects the
behavior of the systems during phase transitions. It is known that
the effect of quenched impurities manifests itself in the form of
random disturbances of the local temperature in ferro and
antiferromagnetic systems in the absence of an external magnetic
field or in the form of random magnetic fields in antiferromagnetic
systems in a uniform magnetic field.

The investigations performed showed that in the first case the
presence of quenched nonmagnetic impurity atoms changes the
properties of only anisotropic Ising like magnets in phase
transitions in which the heat capacity in the "purely" uniform state
diverges at a critical temperature. In the opposite case of magnets
described by the XY model or by the isotropic Heisenberg model, the
presence of an impurity does not affect their behavior during phase
transitions. The critical properties of the disordered Ising model
has recently been studied in numerous works (see references in
\cite{1}). For the dilute Ising-like systems, good agreement was
obtained between the theoretical calculations and the results of
experiments and computer simulation by the Monte Carlo method. An
opposite situation is observed for the magnetic systems with a
disorder of the random magnetic field type. In spite of numerous
studies, which began since 1975 when this type of disorder was
described for the first time \cite{2}, only scarce reliable
information on the behavior of these systems exists at present
\cite{3}. In particular, the nature of the phase transition in the
three-dimensional Ising model with random fields still remains
unexplained, and the theoretical results obtained contradict the
experimental data. The comparison of theoretical predictions with
the results of experimental studies is hindered by the difficulties
of achieving the equilibrium state in such systems because of their
anomalously slow relaxation properties. According to some data, this
is a first-order transition \cite{4,5} up to very low values of
random fields; according to other, this is a second-order transition
\cite{6,7}.

For describing the influence of random fields on the behavior of
magnetic systems, two qualitatively equivalent models, namely, the
ferromagnetic random-field Ising model (RFIM) \cite{8,9} and the
disordered antiferromagnetic Ising model in an external uniform
field (DAFF) \cite{10} are used. The real magnetic systems with the
effects of random fields are antiferromagnets with quenched
impurities of nonmagnetic atoms; in the behavior of these systems,
effects of a ferromagnetic interaction of next-nearest atoms
manifest themselves along with the antiferromagnetic interaction of
nearestneighbor atoms. The structure of an antiferromagnet can be
represented as several ferromagnetic sublattices inserted into each
other in such a way that the total magnetization of the
antiferromagnet remains equal to zero despite the fact that at a
temperature lower than the N\'{e}el point there occurs a magnetic
ordering within each ferromagnetic sublattice. As examples of
two-sublattice antiferromagnets, $\mathop{\mathrm{NiO}}$,
$\mathop{\mathrm{MnO}}$, $\mathop{\mathrm{Fe_2O_3}}$,
$\mathop{\mathrm{MnF_2}}$, and some other can be mentioned. As the
examples of the realization of disordered systems with random
magnetic fields, uniaxial Ising-like antiferromagnets such as
$\mathop{\mathrm{MnF_2}}$ and $\mathop{\mathrm{FeF_2}}$ with
impurities of $\mathop{\mathrm{Zn}}$ atoms in an external magnetic
field can be taken \cite{11}.

In \cite{12,13}, we for the first time showed, using Monte Carlo
computer simulation of the thermodynamic behavior of the disordered
antiferromagnetic Ising model with effects of random magnetic
fields, that, in weakly disordered systems with a spin concentration
higher than the threshold for impurity percolation, a second-order
phase transition from the paramagnetic into the antiferromagnetic
state is realized. In strongly disordered systems with a spin
concentration lower than this threshold value, a first-order
transition from the paramagnetic into a mixed state, which is
characterized by a complex domain structure consisting of
antiferromagnetic and ferromagnetic domains separated by regions of
a spin–glass phase, occurs in the system. It was shown that, with a
reduction in the spin concentrations and an increase in the
magnitude of the external magnetic field, a decrease in the number
and sizes of antiferromagnetic domains occurs in the system, and an
increase in the number and sizes of ferromagnetic domains along with
a reduction of the relative volume of the spin-glass phase is
observed. It is shown that in this region of spin concentrations the
effects of random magnetic fields lead to a change from an
antiferromagnetic ground state to a spin-glass state.

In this work, we consider the same antiferromagnetic Ising model
(with a spin concentration $p = 0.5$ corresponding to the region of
strong structural disordering) as in \cite{12,13}. The aim is to
obtain an additional confirmation of the existence in such systems
of a spin-glass ground state and of a complex domain structure by
the realization and application, for its numerical study, of the
algorithm of the parallel-tempering method which was developed
specially for studying the thermodynamics of spin glasses. To
compare the results obtained, we also investigated the model at
$p=0.9$ corresponding to the region of weak disorder.

\section{Model\label{sec:2}}
The disordered two-sublattice antiferromagnetic Ising model was
defined as a system of spins with a concentration p connected with
$N = pL^3$ sites of a cubic lattice with periodic boundary
conditions. The Hamiltonian of the model under consideration has the
form
\begin{equation}
 {\cal H} = J_1 \sum_{i, j} p_i p_j \sigma_i\sigma_j  + J_2 \sum_{i, k} p_i p_k \sigma_i\sigma_k  +  \mu h \sum_i p_i \sigma_i ,
\end{equation}
where $\sigma_{i}=\pm 1$, $\mu$ is the Bohr magneton, $J_1 = 1$ and
$J_2 = - 1/2 $ characterize the antiferromagnetic interaction of
spins with nearest neighbors and their ferromagnetic interaction
with next-nearest neighbors, respectively; and $h$ is the intensity
of a uniform magnetic field. The random variables $p_i$ and $p_j$
are described by the distribution function
\begin{equation}
P(p_i)=p\delta(p_i-1)+(1-p)\delta(p_i)
\end{equation}
and characterize the quenched nonmagnetic impurity atoms distributed
over the lattice sites (empty sites).

The simulation of statistical properties of this model made it
possible to determine thermodynamic values such as the total
magnetization
\begin{equation}
M=\frac{1}{pL^3}\left[\left<\sum_{i}{p_i\sigma_{i}}\right>\right],
\label{M}
\end{equation}
the "staggered" magnetization $M_{stg} = M_1 -  M_2$ ($M_1$ and
$M_2$ -- are the magnetizations of the sublattices), and the
spin-glass order parameter
\begin{equation}
q_{\alpha,\beta}=\frac{1}{pL^3}\left[\sum_{i}\left<p_i\sigma_i\right>_{(\alpha)}\left<p_i\sigma_i\right>_{(\beta)}\right]\,,
\label{Q}
\end{equation}
where the exponents $\alpha$ and $\beta$ characterize different
replicas of the disordered system that are simulated simultaneously
at one and the same temperature and have different initial
configurations. In expressions (\ref{M}) and (\ref{Q}), the angular
brackets designate statistical averaging realized for each impurity
configuration of the system, and the brackets mean averaging over
different impurity configurations.

The quantities $M$, $M_{\text{stg}}$, and $q_{\alpha,\beta}$, â
characterize different types of magnetic ordering of a strongly
disordered system that can appear in it in the low-temperature
phase. Along with these magnetic thermodynamic quantities,
measurements of the heat capacity as a thermal characteristic of
occurring phase transformations in the system were realized.

\section{Parallel-tempering method\label{sec:3}}
As is known, the spin-glass state is characterized by the presence
of a large number of metastable energy states separated by potential
barriers. The number of metastable states exponentially grows with
increasing number of spins, which strongly complicates the numerical
simulation of such systems. In spin glasses, there is a problem of
reaching an equilibrium state because of the existence of high
energy barriers which separate local energy minima. At sufficiently
low temperatures, the system can never leave a local energy minimum
even if the corresponding state is globally unstable. This feature
makes impossible obtaining physical characteristics for the magnetic
materials which contain, as in our case, a spin-glass phase when
using standard Monte Carlo algorithms.
\begin{wrapfigure}[12]{r}{7cm}   
\vspace*{-2mm}\includegraphics[width=0.45\textwidth]{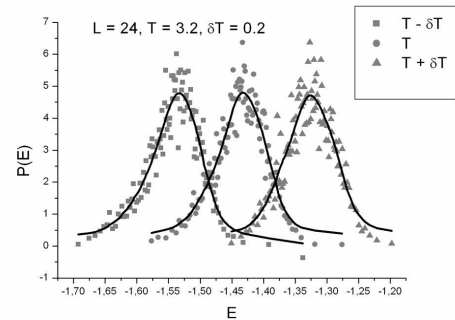}\\[2mm]
\vspace*{-20mm}\\
\caption{ \label{fig:1} Overlap of energy distributions for adjacent
replicas. }
\end{wrapfigure}
There arises a need for an improvement or a modification of the
algorithms used for simulation. One of the algorithms which made it
possible to solve this problem became the parallel-temperature
algorithm \cite{14}.

The parallel-tempering method is an extension of the usual
Metropolis algorithm. The optimization consists in the addition to
the Metropolis algorithm of a second Markov chain in the temperature
parameter $\beta=J_1/T$. The new distribution law is in this case
written as
\begin{equation}
P(\{\sigma\}, \{\beta_{\alpha}\}) \sim
\exp(-\beta_{\alpha}H(\{\sigma\}) + g_{\alpha}),
\end{equation}
ãäå $g_{\alpha}$ is a certain constant. To each $\beta_{\alpha}$
there corresponds its own $g_{\alpha}$. The probability of changing
the dynamic parameter $\beta_{\alpha}$ will obey an $\exp~( -S )$,
law, where
\begin{equation}
 S = (\beta\,'_{\alpha} - \beta_{\alpha})H(\{\sigma\}) + (g\,'_{\alpha}
 - g_{\alpha}).
\end{equation}
In the case of the simultaneous simulation of the replicas of the
system for each $\beta_{\alpha}$ with $\alpha = 0\dots N$, the
change in the temperature can depend only on the nearest values of
$\alpha$; i.e., we have $\beta\,'_{\alpha} = \beta_{\alpha \pm 1}$.
The probability of a transition between the states that are
determined by adjacent temperatures will be written as
$\exp~(-\Delta S)$, where $\Delta S$ is assigned by the expression
\begin{equation}
\Delta S = S\,' - S = (E_{\beta_{\alpha}} -
E_{\beta_{\alpha+1}})(\beta_{\alpha+1} - \beta_{\alpha}).
\end{equation}
According to this algorithm, the temperature of the system can in
the process of simulation both decrease ("annealing" of the system)
and increase, which makes it possible for the system to surmount
high potential barriers.

The basic criterion for the application of the parallel-tempering
method is the overlap of the energy-distribution functions of the
replicas of the system for adjacent temperatures (Fig.~\ref{fig:1}).
Since the energy variance approaches zero with decreasing
temperature, the selection of equidistant temperatures is not
justified; the interval between adjacent temperatures $\delta T =
T_{\alpha+1} - T_\alpha$  also should decrease. This dependence
leads to additional difficulties upon the simulation of the behavior
of these systems in the region of low temperatures, since the
initial maximum temperature of simulation $T_N$ should be
sufficiently large to surmount the potential barriers which separate
the local energy minima. Meanwhile, the number of replicas of the
system (at different temperatures) considered in simulation must
ensure the mechanism of temperature exchange. Under the conditions
of a finite number of the simulated replicas, which is limited by
the computational resources of the researcher, it is necessary to
solve the problem of an optimum choice of the number of temperatures
to be considered. Theoretically, the most optimum selection of
temperatures, according to \cite{15,16}, is such at which the
probability of a transition of the replica to a new temperature is
constant for the entire set of the replicas:
\begin{equation}
P(E_\alpha, \beta_{\alpha} \to E_{\alpha+1}, \beta_{\alpha+1})  =
\min[1, \exp(-\Delta S)] \simeq \mathop{\mathrm{const}}. \label{P_E}
\end{equation}
As an optimum sequence of temperatures, we can use the geometric
progression $T_{\alpha + 1}/T_\alpha = \mathop{\mathrm{const}}$
with:
\begin{equation}
 T_{\alpha} = T_0R^{\alpha} (\alpha = 0 .. N) ,
 \label{R_N}
\end{equation}
where $R = \sqrt[N]{T_N / T_0}$.

The choice of the temperature sequence in this form ensures the
fulfillment of equality (\ref{P_E}). The optimum value of the
probability of the replica transition to a new temperature upon the
realization of the parallel-tempering algorithm was determined
\cite{17} to be $P \simeq 0.23$.

However, in contrast to the Ising spin glasses, in which the
temperature of transition into the spin-glass state is $T_f/J
\approx 1$, in the dilute antiferromagnetic Ising model with the
effects of random magnetic fields at a spin concentration $p = 0.5$
the temperature of the phase transition into the mixed state is $T_m
\approx 5$ in the units of exchange integral $J_1$. As a result, the
application of the parallel-tempering method to the simulation of
such a system requires the use of a considerably wider temperature
range. Thus, to guarantee obtaining a stable initial equilibrium
state from which it is better to perform simulation, the initial
maximum temperature $T_N$ must be selected to be considerably higher
than the temperature $T_{\text{mix}}$; only after this it is
possible to investigate the phase transition into the mixed state
near $T_{\text{mix}}$ in the vicinity of which there already appears
a set of metastable states with an extremely slow dynamics of
establishing equilibrium. On the other hand, for studying the
asymptotic approach to the ground spin-glass state of the system it
is desirable that the minimum temperature $T_0$ be as close as
possible to $T = 0$. All this imposes new restrictions on the
realization of the parallel-tempering algorithm for studying
disordered antiferromagnets and for the selection of the sequence of
temperatures for the replicas to be simulated.

Note also that for the realization of the condition of the equal
probability of the temperature exchange between the adjacent
replicas it is necessary to have information on the temperature
dependence of the energy of the system $E(T)$; for this reason, the
procedure of finding an optimum set of temperature points requires a
preliminary simulation of the system. As the first approximation, a
sequence of temperatures in the form of a geometric progression
(\ref{R_N}) can be chosen. The subsequent simulation of the system
for the selected temperatures gives an idea about the $E(T)$
dependence, though also only in the first approximation. At the next
step, a new set of temperatures $\beta'_\alpha$ is determined
according to condition (\ref{P_E}) to yield
\begin{eqnarray}
\left(E(\beta'_{\alpha}) - E(\beta'_{\alpha+1})\right)
(\beta'_{\alpha+1} - \beta'_{\alpha}) \simeq
\mathop{\mathrm{const}}. \label{B1}
\end{eqnarray}

The new set of temperatures obtained in this case specifies new
(more optimum) parameters for the simulation of the disordered
system. In turn, the thus obtained replica-temperature distribution
is also an approximation (second-order), which can be used for the
subsequent iterative search for the optimum set of temperatures.

The inconvenience of this approach is the need to, each time anew,
conduct the "complete" simulation of the system at each step of the
search for an optimum sequence of temperatures. Nevertheless, this
approach makes it possible to most effectively use the
parallel-tempering algorithm in conducting numerical experiments on
studying the low-temperature properties of spin-glass states.

\section{Results and discussion\label{sec:4}}
In this work, we consider the antiferromagnetic Ising model with a
spin concentration $p = 0.5$ (corresponding to the region of strong
disordering) and a magnetic field $h = 2$. The simulation was
conducted for a wide set of values of the linear dimensions of the
cubic lattice ($L = 8, 16, 24, 32$, and $40$).

\begin{figure}
\includegraphics[width=0.45\textwidth]{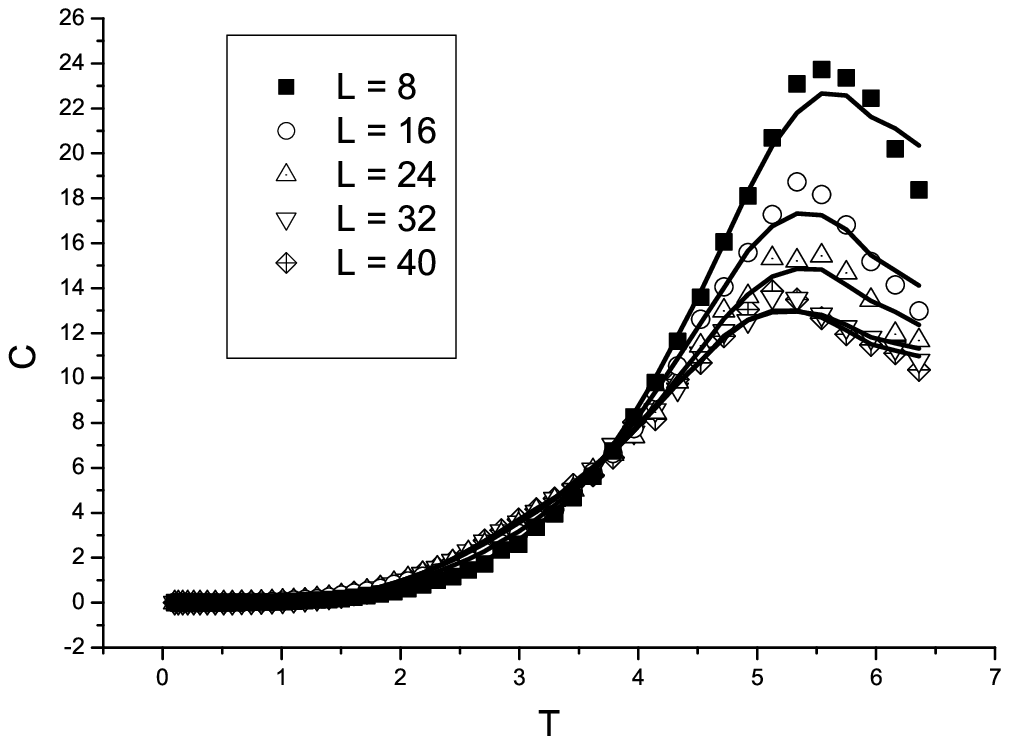} \hfill
\includegraphics[width=0.45\textwidth]{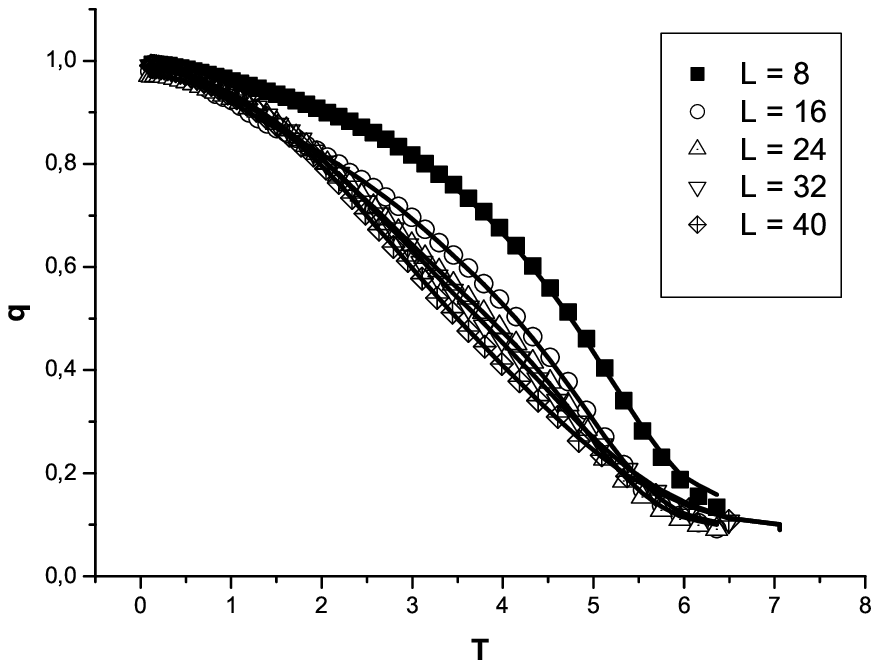} \\
\parbox{0.45\textwidth}{\caption{\label{fig:2}  Temperature dependence of the heat capacity $C$ for
systems with a spin concentration $p = 0.50$ and linear sizes $L =
8, 16, 24, 32$, and $40$.}} \hfill
\parbox{0.45\textwidth}{\caption{\label{fig:3} Temperature dependence of the spin-glass order
parameter $q$ for the systems with a spin concentration $p = 0.50$
and linear sizes $L = 8, 16, 24, 32$, and $40$. }}
\end{figure}

For determining the temperature of the phase transition in the
system and the intervals of existence of different phase states, we
studied the temperature behavior of the heat capacity of the system
(Fig.~\ref{fig:2}) for the lattices of the above-indicated
dimensions. It is evident that in the temperature interval of $T =
4.5–6.5$ there is observed an anomalous increase in the heat
capacity, which indicates the existence of a phase transition in the
system. The size-dependent changes in the behavior of the heat
capacity indicate the suppression of the energy fluctuations in the
system in comparison with the typical second-order phase transitions
into the antiferromagnetic state, and the bend in the temperature
dependence $C(T)$ at $T_{\text{mix}} = 5.13$ for the lattice with
$L_{\text{max}} = 40$ is typical of the phase transition in spin
glasses \cite{3}.

For obtaining equilibrium magnetic characteristics of the system,
the initial states were selected in the paramagnetic phase. This
choice is caused by the fact that, because of the presence of
metastable states near the transition temperature and in the entire
low-temperature phase, there arises a problem of reaching
equilibrium initial configurations. The initial states obtained were
used in the realization of the parallel-tempering method. Within the
framework of this method, the initial set of temperatures was
selected according to formula (\ref{R_N}). The temperature
dependence of the energy $E(T)$ obtained in the simulation was used
for refining the temperatures $\beta_{\alpha}$ according to the
principle of the equal probability of transition between adjacent
temperatures. To achieve the equilibrium state of the system at each
temperature, $10^4$ Monte Carlo steps were used for the relaxation,
with the rejection of half initial configurations.

\begin{figure}
\includegraphics[width=0.45\textwidth]{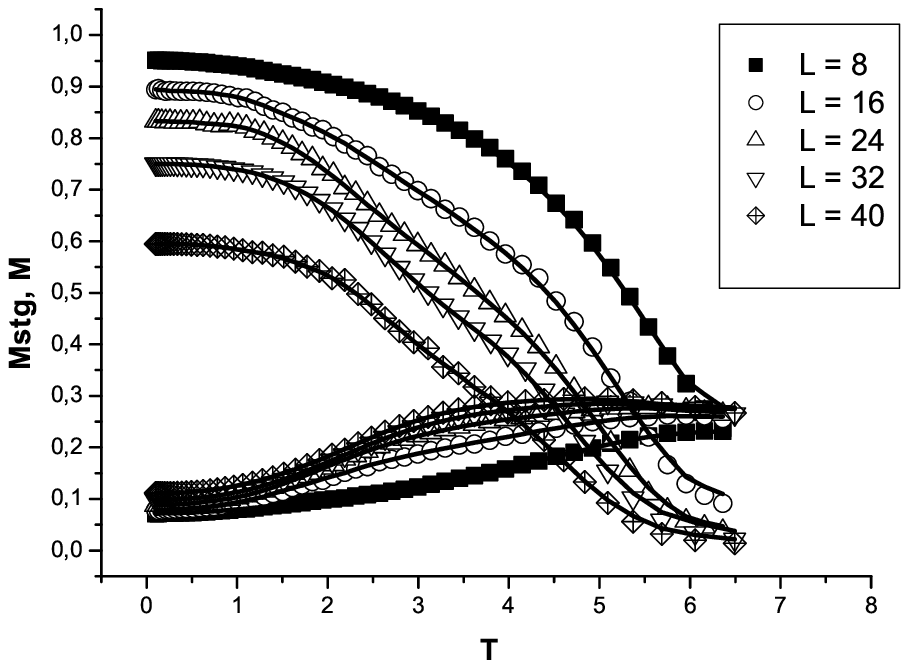} \hfill
\includegraphics[width=0.45\textwidth]{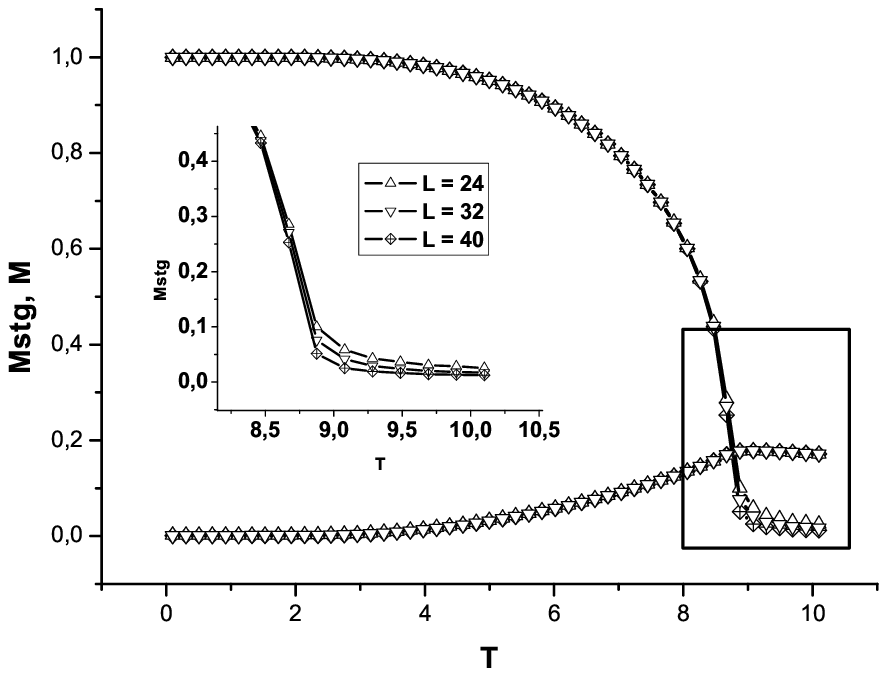} \hspace*{5mm}\\
\parbox[t]{0.45\textwidth}{\caption{\label{fig:4}  Temperature dependence of the "staggered" ($M_{\text{stg}}$,
upper curves) and total ($M$, lower curves) magnetizations for
systems with a spin concentration $p = 0.50$ and linear sizes $L =
8, 16, 24, 32$, and $40$.}} \hfill
\parbox[t]{0.5\textwidth}{\caption{\label{fig:5} Temperature dependence of "staggered" ($M_{\text{stg}}$, upper
curves) and total ($M$, lower curves) magnetizations for systems
with a spin concentration $p = 0.90$ and linear sizes $L = 24, 32$,
and $40$. In the inset, the temperature range of phase transition is
shown for $M_{\text{stg}}$. }}
\end{figure}

Figures~\ref{fig:3}-\ref{fig:4} display the temperature dependences
obtained for the staggered ($M_{\text{stg}}$) and total ($M$)
magnetizations and spin-glass order parameter ($q$) for the lattices
with linear dimensions from $L = 8$ to $L = 40$ averaged over $100$
different impurity configurations. It is seen from the figures that
all the values measured demonstrate a noticeable dependence on the
sizes of the system. The strongest dependence on $L$ is
characteristic of the "staggered" magnetization, which for the
systems of small sizes determines the prevailing magnetic ordering
of antiferromagnetic nature; the spin-glass order parameter in this
case appears as a secondary parameter of ordering and repeats the
temperature dependence of the "staggered" magnetization. The strong
decrease in $M_{\text{stg}}$ with increasing L at noticeably smaller
changes in the spin-glass order parameter indicates the predominance
of spin-glass ordering in the system with $L > 24$ and the
appearance, in the low-temperature phase, of a mixed phase state
consisting of antiferromagnetic and ferromagnetic domains surrounded
by a spin-glass phase. Note that at a temperature $T_{\text{mix}} =
5.13$ the temperature dependence of the total magnetization for the
lattice with $L = 40$ reaches a maximum with a subsequent
characteristic decrease in $M(T)$ in the high-temperature region, as
was observed in \cite{13}. The results obtained indicate that, in
the limit of $L \to \infty$ and $T \to 0$, a spin-glass ground state
is realized in the system.

\begin{wrapfigure}[30]{r}{8cm}   
\vspace*{-23mm}\includegraphics[width=0.45\textwidth]{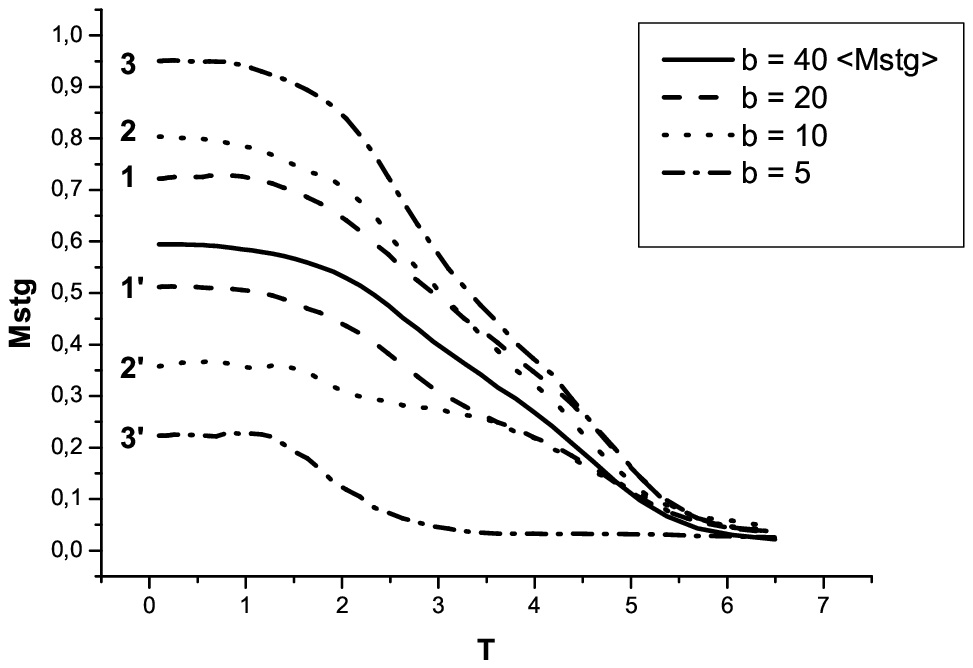}\\[2mm]
\vspace*{-20mm}\\
\caption{ \label{fig:6} Temperature dependence of the local values
of the "staggered" magnetization $M_{\text{stg}}$ for blocks with
sizes $b = 5, 10$, and $20$. }
\vspace*{-2mm}\includegraphics[width=0.45\textwidth]{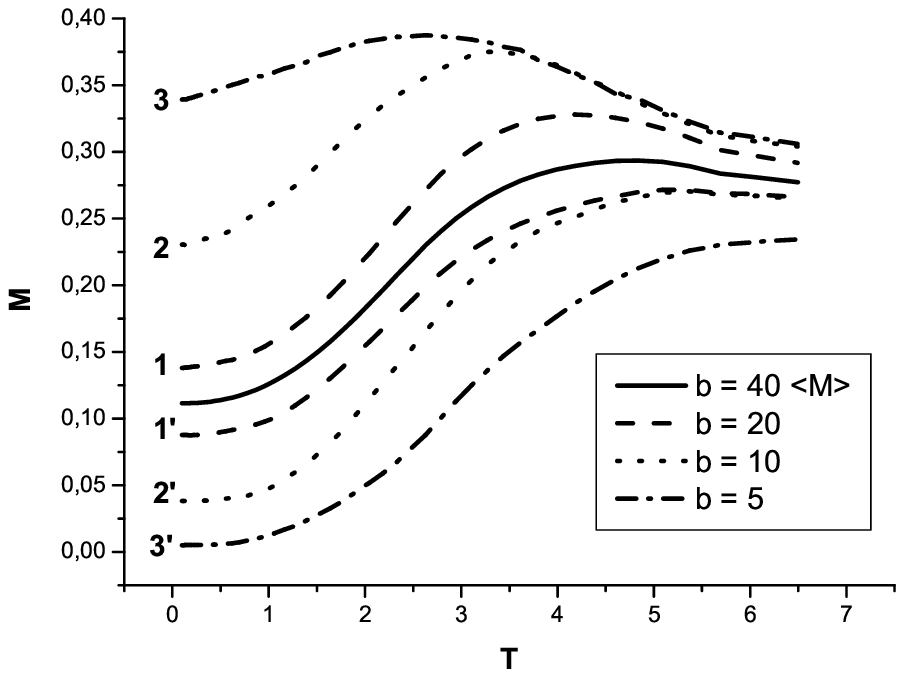}\\[2mm]
\vspace*{-20mm}\\
\caption{ \label{fig:7} Temperature dependence of the local values
of the total magnetization $M$ for blocks with sizes $b = 5, 10$,
and $20$. }
\vspace*{-2mm}\includegraphics[width=0.45\textwidth]{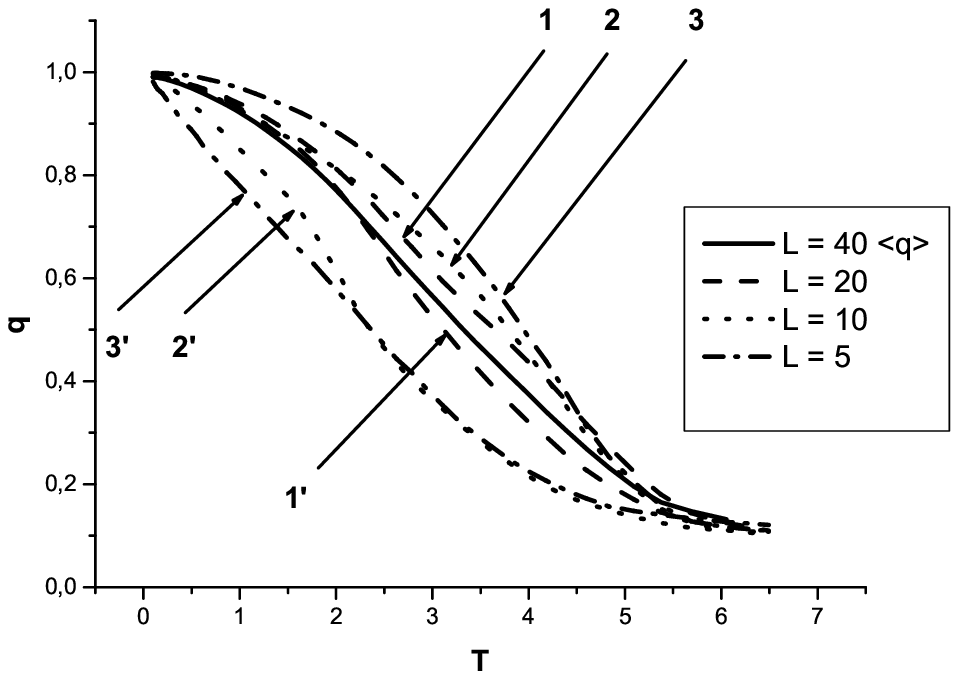}\\[2mm]
\vspace*{-20mm}\\
\caption{ \label{fig:8} Temperature dependence of the local values
of the spin-glass order parameter $q$ for blocks with sizes $b = 5,
10$, and $20$. }
\end{wrapfigure}

To compare the observed features of the behavior of the magnetic
characteristics of the model in the region of strong structural
disorder with the behavior of analogous characteristics in the
region of weak dilution, we carried out an analogous study for the
samples with a spin concentration $p = 0.9$. Figure~\ref{fig:5}
displays the temperature dependences of the "staggered"
($M_{\text{stg}}$) and total ($M$) magnetizations on the linear
dimensions of the lattice $L = 24, 32$, and $40$. It is evident that
no size dependence of these characteristics is observed in the
low-temperature phase, except for the temperature range close to the
critical temperature $T_c = 8.73$. The temperature behavior of
$M_{\text{stg}}$ clearly indicates the existence of an
antiferromagnetic ordering in the system with an the
antiferromagnetic ground state under above conditions.

To confirm the existence of a domain structure in the mixed phase
state of the strongly disordered antiferromagnet with random fields,
we realized, within the framework of the statistical
parallel-tempering method, a study of the temperature dependence of
the local values of magnetic characteristics for cubic blocks of
size $b = 5, 10$, and $20$ into which the lattice under
consideration (with a maximum size $L = 40$) was divided. These
dependences are shown in Figs.~\ref{fig:6}-–\ref{fig:8}; "solid"
curves correspond to the dependence of the average values of
$M_{\text{stg}}(T)$, $M(T)$, and $q(T)$ for the entire lattice with
$L = 40$; curves $1$ and $1'$, $2$ and $2'$, and $3$ and $3'$
correspond to minimum and maximum values of these values for the
blocks with $b = 5, 10$, and $20$, respectively. An analysis of
these figures and the data on the entire totality of blocks shows
that as the temperature decreases, the sizes of typical
antiferromagnetic domains decrease from $l_a \simeq 20$ to $la
\simeq 10$, and the sizes of ferromagnetic domains, from $l_f \simeq
10$ to $l_f \simeq 5$ with an increase in the volume of the
spin-glass phase, until at $T = 0$ there is realized a spin-glass
ground state.
\section*{Conclusions}
Thus, by applying the numerical parallel-tempering method for
studying the low-temperature behavior of the three-dimensional
disordered antiferromagnetic model with random magnetic fields, it
was clearly shown that in weakly disordered systems there is
realized an antiferromagnetic ordered state, whereas in the region
of strong structural disorder the effects of random magnetic fields
lead to the realization of a new phase state in the system, which is
characterized by a complex domain structure consisting of
antiferromagnetic and ferromagnetic domains separated by regions of
a spin-glass phase with the realization of a spin-glass ground
state.
\section*{Acknowledgements}
This work was supported in part by the Russian Foundation for Basic
Research, project nos. 04-02- 17524 and 04-02-39000.


\begin{thebibliography}{99}
\bibitem{1} R. Folk, Yu. Holovatch, and T. Yavorskii, Critical Exponents of a Three-Dimensional Weakly Diluted Quenched Ising Model,
Usp. Fiz. Nauk \textbf{173} (2), 175--200 (2003) [Phys.-Usp. \textbf{46}
(2), 169--191 (2003)].
%
\bibitem{2} Y. Imry and S.-K. Ma, Random-Field Instability of the Ordered State of Continuous Symmetry,
Phys. Rev. Lett. \textbf{35} (21), 1399--1401 (1975).
%
\bibitem{3} V. S. Dotsenko, Critical Phenomena and Quenched Disorder,
Usp. Fiz. Nauk \textbf{165} (5), 481--528 (1995) [Phys.-Usp. \textbf{38} (5), 457--496 (1995)].
%
\bibitem{4} A.P. Young and M. Nauenberg, Quasicritical Behavior and First-Order Transition in the $d = 3$ Random-Field Ising Model,
Phys. Rev. Lett. \textbf{54} (22), 2429--2432 (1985).
%
\bibitem{5} H. Rieger and A. P. Young, Critical Exponents of the Three-Dimensional Random Field Ising Model,
J. Phys. A \textbf{26} (20), 5279--5284 (1993).
%
\bibitem{6} A. T. Ogielski and D. A. Huse, Critical Behaviour of the Three-Dimensional Dilute Ising Antiferromagnet in a Field,
Phys. Rev. Lett. \textbf{56} (12), 1298--1301 (1986).
%
\bibitem{7} A. T. Ogielski, Integer Optimization and Zero-Temperature Fixed Point in Ising Random-Field Systems,
Phys. Rev. Lett. \textbf{57} (10), 1251--1254 (1986).
%
\bibitem{8} J. Cardy, Random-Field Effects in Site-Disordered Ising Antiferromagnets,
Phys. Rev. B: Condens. Matter \textbf{29} (1), 505--507 (1984).
%
\bibitem{9} D.P. Belanger and A. P. Young, The Random Field Ising Model,
J. Magn. Magn. Mater. \textbf{100}, 272--291 (1991).
%
\bibitem{10} G. S. Grest, C.M. Soukoulis, and K. Levin, Comparative Monte Carlo and Mean-Field Studies of Random-Field Ising Systems,
Phys. Rev. B: Condens. Matter \textbf{33} (11), 7659--7674 (1986).
%
\bibitem{11} F. Ye, L. Zhou, S. Larochelle, et al., Order Parameter Criticality of the $d = 3$ Random-Field Ising Antiferromagnet $\mathop{\mathrm{Fe_{0.85}Zn_{0.15}F_2}}$,
Phys. Rev. Lett. \textbf{89}, 157202--157205 (2002).
%
\bibitem{12} V. V. Prudnikov, O. N. Markov, and E. V. Osintsev, Peculiarities of Phase Transformations in a Random-Field Ising Antiferromagnet,
Zh. Eksp. Teor. Fiz. \textbf{116} (3), 953--961 (1999).
%
\bibitem{13} V. V. Prudnikov and V. N. Borodikhin, Monte Carlo Simulation of a Random-Field Ising Antiferromagnet,
Zh. Eksp. Teor. Fiz. \textbf{128} (2), 337--343 (2005)
[J. Exp. Theor. Phys. \textbf{101} (2), 294--298 (2005)].
%
\bibitem{14} J. J. Moreno, H. G. Katzgraber, and A. K. Hartmann, Finding Low-Temperature States with Parallel Tempering, Simulated Annealing and Simple Monte Carlo,
Int. J. Mod. Phys. A \textbf{14} (3), 285--298 (2003).
%
\bibitem{15} D. A. Kofke, On the Acceptance Probability of Replica-Exchange Monte Carlo Trials,
J. Chem. Phys. \textbf{117} (15), 6911--6914 (2002).
%
\bibitem{16} D. A. Kofke, Erratum: On the Acceptance Probability of Replica-Exchange Monte Carlo Trials,
J. Chem. Phys. \textbf{120} (22), 10852--10852 (2004).
%
\bibitem{17} A. Kone and D. A. Kofke, Selection of Temperature Intervals for Parallel-Tempering Simulations,
J. Chem. Phys. \textbf{122}, 206101--206102 (2005).
%
\end{thebibliography}
\end{document}